\documentclass[aps,showpacs,superscriptaddress,nofootinbib,amsmath,amssymb]{revtex4}

 \usepackage{graphicx}
\usepackage{dcolumn}
\usepackage{bm}
\usepackage{lipsum}
\usepackage{adjustbox}
\usepackage{multirow}
\begin{document}
\title{Nuclear dependence of $R=\frac{\sigma_L}{\sigma_T}$ and Callan-Gross relation in nuclei}
\author{F. Zaidi}
\author{H. Haider}
\author{M. Sajjad Athar\footnote{Corresponding author: sajathar@gmail.com}}
\author{S. K. Singh}
\affiliation{Department of Physics, Aligarh Muslim University, Aligarh - 202002, India}
\author{I. \surname{Ruiz Simo}}
\affiliation{Departamento de F\'{\i}sica At\'omica, Molecular y Nuclear,
and Instituto de F\'{\i}sica Te\'orica y Computacional Carlos I,
Universidad de Granada, Granada 18071, Spain}
\begin{abstract}
The electromagnetic nuclear structure functions $F_{1A} (x,Q^2)$, $F_{2A} (x,Q^2)$ and 
 $F_{LA} (x,Q^2)$ have been calculated using a 
 microscopic model of nucleus to study the nuclear medium effects on the ratio 
  $R_A(x,Q^2)=\frac{\sigma_{LA} (x,Q^2)}{\sigma_{TA} (x,Q^2)} = \frac{F_{LA} (x,Q^2)}{2xF_{1A} (x,Q^2)}$
and the Callan-Gross relation(CGR) in nuclei. The nuclear medium effects
due to the Fermi motion, binding energy, nucleon correlations, mesonic contribution and shadowing have been taken into account.
 The theoretical results for the nuclear dependence of $R_{A} (x,Q^2)$ and its impact on CGR have been presented and compared with the
 available experimental data on the various nuclear targets.
 The predictions have been made for $R_{A} (x,Q^2)$ in the kinematic region of $x$ and $Q^2$ for some nuclei relevant for the future experiments to be performed at the JLab.
  \end{abstract}
\maketitle
 \section{Introduction} 
 A better theoretical understanding of the nuclear medium effects in the deep inelastic scattering(DIS) region 
  in the electromagnetic(EM) and weak interaction induced processes has been emphasized~\cite{Bodek:2015mwa, Guzey:2012yk, Solvignon:2009it, Kovarik:2010uv, Haider:2016zrk}
  in view of the present DIS experiments being performed on various nuclear targets
 using  the electron beams at the JLab~\cite{Mamyan:2012th, jlab, jlab_current, Covrig} and the neutrino/antineutrino beams at the Fermilab~\cite{Fermilab}. A dedicated experiment at the JLab 
 to study the nuclear medium effects in the kinematic
 region of $Q^2(1<Q^2<5~GeV^2)$ and $x(0.1<x<0.6)$ for the electron induced DIS process on hydrogen, deuterium, carbon, copper and gold targets has been proposed~\cite{Covrig}.
 Nuclear medium effects are also being studied in the
 $\nu_l/\bar\nu_l-$nucleus scattering in the nuclear targets like carbon, iron and lead in the MINER$\nu$A experiment at the Fermilab~\cite{Mousseau:2016snl}.
 Generally, the experimental results of the cross section for DIS processes induced by the
 charged leptons and the neutrinos/antineutrinos on the nucleons and the nuclear targets are interpreted in terms of the  
structure functions. 
In the case of EM DIS processes induced by the leptons on the nucleons, the cross sections are
described in terms of the two nucleon structure functions $F_{1N}  (x,Q^2)$ and $F_{2N}  (x,Q^2)$,
where $x(=\frac{Q^2}{2 M_N \nu})$ is the Bjorken scaling variable, $M_N$ is the mass of target nucleon, $\nu(=E-E^\prime)$
and $Q^2(=4EE^\prime sin^2\left({\theta \over 2}\right))$ are the energy and four momentum transfer square to the hadronic system. 
The structure function $F_{1N}(x,Q^2)$ describes the contribution of the transverse component of the virtual photon to the DIS cross sections
while the structure function $F_{2N}(x,Q^2)$ describes a linear combination of the longitudinal and transverse
components. Alternately, the DIS cross section is also described in terms of the transverse structure function $F_{TN}(x,Q^2)$ 
and the longitudinal structure function $F_{LN}(x,Q^2)$ defined as
\begin{equation}
F_{TN} (x,Q^2)= 2x F_{1N}(x,Q^2)\;;\;\;\;\;
 F_{LN} (x,Q^2)=\left(1+\frac{4 M_N^2 x^2}{Q^2} \right) F_{2N} (x,Q^2) - 2 x  F_{1N} (x,Q^2).
\end{equation}
The transverse and longitudinal cross sections are then expressed as
\begin{equation}
     \sigma_{TN,LN}(x,Q^2) = \left(\frac{4\pi^2\alpha}{2x\nu(1-x)M_N} \right) F_{TN,LN} (x,Q^2). 
\end{equation}

 The ratio of the purely longitudinal to transverse cross sections, $R_N(x,Q^2)$ is defined as
\begin{eqnarray}\label{rlnr2n}
 R_N(x,Q^2)&=&\frac{\sigma_{LN}(x,Q^2)}{\sigma_{TN}(x,Q^2)}={F_{LN} (x,Q^2) \over 2 x F_{1N} (x,Q^2)}=\left(1+\frac{4 M_N^2 x^2}{Q^2} \right) R_{2N} (x,Q^2) - 1;~R_{2N} (x,Q^2)= {F_{2N} (x,Q^2) \over 2 x F_{1N} (x,Q^2)}.
\end{eqnarray}
  In the kinematic region of Bjorken scaling($Q^2 \to \infty,~\nu \to \infty$ such that $x={Q^2 \over 2M_N\nu} \to$constant), all
the nucleon structure functions scale i.e. $F_{iN} (x,Q^2) \rightarrow F_{iN} (x)\;\;(i=1,2,L) $. In this kinematic region, the
structure functions $F_{1N} (x) $ and $F_{2N} (x)$ calculated in the quark-parton model satisfy the Callan-Gross relation given by~\cite{Callan:1969uq}:
\begin{eqnarray}
 \label{cgr}
   F_{2N} (x) &=& 2 x F_{1N} (x)\\
   \label{rr}
\mbox{implying}\;\;\; R_{2N}(x,Q^2) &\rightarrow& 1,\;\;{\mbox{and}}\;\;\;R_N(x,Q^2) \rightarrow 0\;\;\mbox{in the limit of $Q^2$} \rightarrow \infty.
\end{eqnarray}

Therefore, in the kinematic limit of the Bjorken scaling, the EM DIS data on the scattering of the electrons from the proton targets are analyzed in terms of 
 only one structure function $F_{2N} (x)$. An explicit evaluation of $F_{2N} (x)$ in the quark parton model gives~\cite{cooper_sarkar}: 
 \begin{equation}
\label{eq:f22xf1}
F_{2N} (x) = 2 x F_{1N} (x) = x \sum_i e_i^2 \left( f_i(x) + \bar f_i(x) \right),
\end{equation}
where $f_i(x)$ and $\bar f_i(x)$ are the quark and antiquark parton distribution functions(PDFs) which describe the probability of finding a quark/antiquark of
flavor $i$ carrying a momentum fraction $x$ of the nucleon's momentum. $e_i^2$ is the square of the charge corresponding to the quark/antiquark of flavor $i$.

 As we move away from the kinematic region of the validity of Bjorken scaling 
 towards the region of smaller $Q^2$ and $\nu$, the description 
 of structure function becomes more complex and the effects due to the target mass 
  correction(TMC) and the higher twists(HT)
 as well as other non perturbative
QCD effects arising due to the quark-quark and quark-gluon interactions are expected to give rise to $Q^2$ dependent contribution to the structure 
 functions which violate scaling.
 Theoretical studies show that the corrections to the 
 nucleon structure functions due to these effects decrease as ${1 \over Q^2}$, and therefore become important at small and moderate
 $Q^2$~\cite{Melnitchouk:2005zr, Schienbein:2007gr,Miramontes:1989ni, Castorina:1984wd, Melnitchouk:2006je}. 
 These contributions may be different for $F_{1N}(x,Q^2)$ and $F_{2N}(x,Q^2)$ leading to the $Q^2$ dependent corrections in CGR given by Eqs.(\ref{cgr}) and (\ref{rr}). 
There exist some phenomenological attempts to study the deviation of $R_N(x,Q^2)$ from its Bjorken limit 
 by studying the $Q^2$ dependence of $F_{LN}(x,Q^2)$ in the region of smaller and moderate $Q^2$~\cite{Whitlow:1991uw, Bodek:2010km, Christy:2007ve, 
 Abe:1998ym, Dasu:1988ms}.
 These phenomenological studies describe the available experimental results on 
 $R_N(x,Q^2)$~\cite{Whitlow:1991uw, Bodek:2010km, Christy:2007ve, 
 Abe:1998ym, Dasu:1988ms, Tvaskis:2006tv, Monaghan:2012et, Liang:2004tj, Benvenuti:1989rh}.
 The most utilized parameterization of $R_N (x,Q^2)$ used in many experimental analyses is by Whitlow et al.~\cite{Whitlow:1991uw}.

In the case of nuclear targets the EM DIS cross sections are similarly analyzed in terms of nuclear structure function $F_{2A} (x,Q^2)$
assuming the validity of CGR at the nuclear level.
A comparative study of nuclear structure function $F_{2A}(x,Q^2)$ with the free nucleon structure function $F_{2N}(x,Q^2)$ 
 led to the discovery of the EMC effect~\cite{Aubert:1983xm, Bodek:1983qn, Bodek:1983qn}. The nuclear medium effects arising due to the  
 Fermi motion, binding energy, nucleon
 correlations, shadowing, etc. in understanding the EMC effect, in the various regions of $x$ has been extensively studied
 in the last 35 years~\cite{Malace:2014uea, Geesaman:1995yd, Hen:2013oha}.
 
However, there have been very few theoretical attempts to make a comparative study of the nuclear medium effects in $F_{LA}(x,Q^2)$, $F_{1A}(x,Q^2)$ and $F_{2A}(x,Q^2)$
 and understand their implications on the modifications of $R_{A}(x,Q^2)$, $R_{2A}(x,Q^2)$ and CGR in nuclei.
 It has been argued that various nuclear medium effects arising due to the Fermi motion of nucleons, mesons, gluons distributions
and higher twist effects in nuclei would contribute differently in the longitudinal and transverse structure functions in the
 various regions of $x$ and $Q^2$ and induce nuclear dependence on 
$R_{A}(x,Q^2)$, $R_{2A}(x,Q^2)$ and CGR in nuclei~\cite{Castorina:2002zf, Miller:2000ta, Kulagin:2004ie,Ericson:2002ep,Armesto:2010tg}.
 Nuclear dependence of $R_{A}(x,Q^2)$ has been the topic of investigation in some experiments~\cite{Dasu:1988ms, Amaudruz:1992wn, Arneodo:1996ru, Gomez:1993ri} but no 
significant nuclear dependence has been reported. However, a reanalysis of these experiments shows a nontrivial nuclear 
dependence of $R_{A}(x,Q^2)$ and its implications on the extraction of the EMC ratio $F_{2A}(x,Q^2) \over F_{2D}(x,Q^2)$ have been studied~\cite{Guzey:2012yk, Solvignon:2009it}. 
 In heavier nuclei, the Coulomb corrections to $R_{A}(x,Q^2)$  due to 
 the initial and final electrons moving in the electrostatic potential V(r) of the nucleus have also been considered and are found  
 to be small and relevant only for DIS of the low energy electrons from the high Z nuclear targets~\cite{Solvignon:2009it}. 
  The recent experimental measurements on the EM nuclear structure functions reported from the JLab on various nuclei in the kinematic region of
 $Q^2(1<Q^2<5)~GeV^2$ and $x(0.1<x<1)$ also show that the nuclear medium effects are different 
 for  $F_{1A}  (x,Q^2)$,  $F_{2A}  (x,Q^2)$ and  $F_{LA}  (x,Q^2)$~\cite{Mamyan:2012th} and could modify the CGR in nuclei. 

 In view of these experimental results a comparative theoretical study of the nuclear structure functions $F_{iA} (x,Q^2)(i=1,2,L)$ for the electromagnetic processes 
 and its effect on $R_{A}(x,Q^2)$, $R_{2A}(x,Q^2)$ and CGR in the nuclear medium in the various regions of $x$ and $Q^2$ is highly desirable. 
 A comparison of the theoretical results 
 with the present and future experimental data
 from the JLab~\cite{Mamyan:2012th,jlab,jlab_current,Covrig} will lead to a better 
 understanding of the nuclear medium effects in EM structure functions.

In this work, we have studied the nuclear medium effects in the structure functions $F_{iA} (x,Q^2)$ ($i=1,2,L$) and nuclear dependence of $R_{A}(x,Q^2)$    
 in the regions of $Q^2$ and $x$ relevant for the present and future 
experiments at the JLab in $^{12}C$, $^{27}Al$, $^{56}Fe$, $^{63}Cu$ and $^{197}Au$, assuming 
  the $Q^2$ and $x$ dependent phenomenological corrections to $R_N(x,Q^2)$ for the free nucleon case given by Whitlow et al.~\cite{Whitlow:1991uw}.
 The impact of the nuclear dependence of $R_A(x,Q^2)$ on the CGR and its dependence on $Q^2$ has also been studied.
 In section \ref{sec_formalism}, the formalism for calculating the electromagnetic
structure functions and the ratios $R_A(x,Q^2)$ and $R_{2A}(x,Q^2)$ in the nuclear 
medium is given in brief. In section \ref{sec_results}, 
 the numerical results are presented
which are summarized in section \ref{sec_summary}. 
\section{Formalism}\label{sec_formalism}
In a nucleus, the charged lepton interacts with the nucleons which are moving
with some momenta constrained by the Fermi momenta and Pauli blocking, 
 and the nucleus is assumed to be at rest. Therefore, the free
 nucleon quark and antiquark PDFs should be convoluted with the momentum distribution of nucleons.
 In addition, there are binding
energy corrections. Furthermore, the target nucleon being strongly interacting particle interacts with the other   
 nucleons in the nucleus leading to the nucleon correlations effects. We have taken these effects into account by using a field theoretical model which starts 
 with the Lehmann's representation for the relativistic nucleon propagator and the nuclear many body theory is used to calculate it for
 an interacting Fermi sea in the nuclear matter. A local density approximation is then applied to obtain the results for 
 a finite nucleus. This technique results into the use of a relativistic nucleon 
spectral function that describes the energy and momentum distributions~\cite{FernandezdeCordoba:1991wf}. 
 All the information like Fermi motion, binding energy and the nucleon correlations are contained in the spectral function. 
 Moreover, we have considered the contributions of the pion and rho mesons 
in a many body field theoretical approach 
based on Refs.~\cite{Marco:1995vb,GarciaRecio:1994cn}. 
 The free meson propagator is replaced by a dressed one as these mesons interact with the nucleons in the
nucleus through the strong interaction.
We have earlier applied this model to study the nuclear medium effects
in the electromagnetic and weak processes~\cite{Haider:2016zrk,Haider:2015}, 
as well as proton induced Drell-Yan processes~\cite{Haider:2016tev} on the nuclear targets.

In the case of a nuclear target, the expression for the differential scattering cross section is given by
\begin{equation}\label{eA}
\frac{d^2 \sigma_A}{d\Omega dE^{\prime}} =~\frac{\alpha^2}{q^4} \; \frac{|\bf k'|}{|\bf k|} \;L_{\mu \nu} \; W_A^{\mu \nu},
\end{equation}
where $\alpha$ is the fine structure constant, $L_{\mu\nu}=2 \left( k_\mu k'_\nu + k'_\mu k_\nu - g_{\mu\nu} k \cdot k' \right)$ 
is the leptonic tensor and $W_A^{\mu \nu}$ is the nuclear hadronic tensor which is expressed in terms of nuclear structure functions $W_{iA} (\nu,Q^2)(i=1,2)$ as
\small{
\begin{equation}
 W^{\mu\nu}_A
= W_{1A} (\nu,Q^2)
  \left( { q^\mu q^\nu \over q^2} - g^{\mu\nu} \right)
+ { W_{2A} (\nu ,Q^2) \over M_A^2 }
  \left( p_A^\mu - { p_A\cdot q \over q^2 } q^\mu \right)
  \left( p_A^\nu - { p_A\cdot q \over q^2 } q^\nu \right),
\label{eq:Wmunu_nucleus}
\end{equation}
}
where $M_A$ is the mass and $p_A$ is the four momentum of the target nucleus. 

The differential scattering cross section may also be written in terms of the 
probability per unit time ($\Gamma$) of finding a charged lepton interacting with a target nucleon given by~\cite{Haider:2016zrk}:
\begin{equation}\label{defxsec1}
 d\sigma=\Gamma~dt~dS=\Gamma~\frac{dt}{dl}~dl~dS=\frac{\Gamma}{ v}dV=\Gamma\frac{E({\bf k})}{|{\bf k}|}dV=\frac{-2m_l}{\mid {\bf k} \mid} Im \Sigma (k) dV,
\end{equation}
\begin{figure}
 \includegraphics[height=3.5 cm, width=8.0 cm]{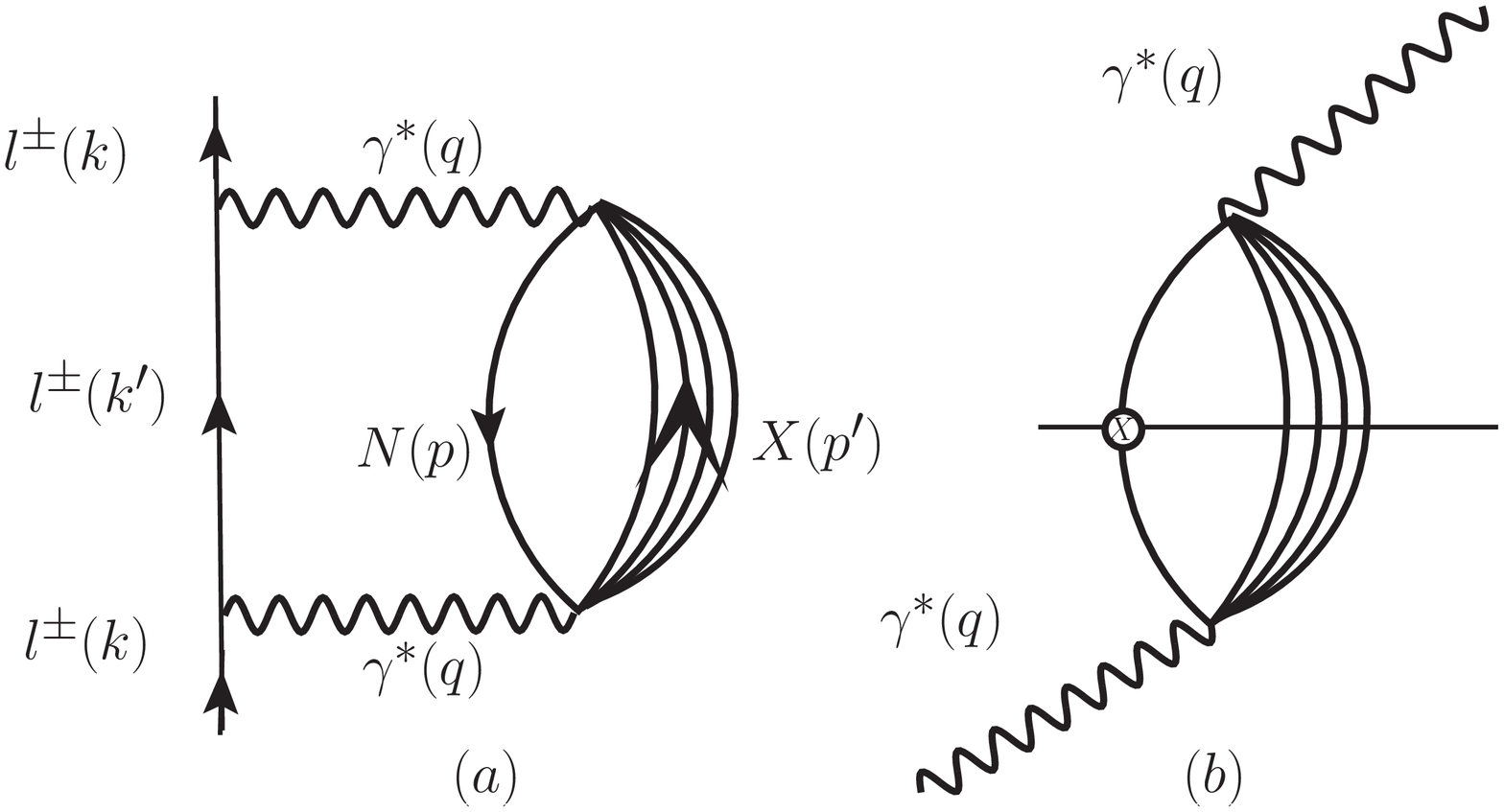}
 \caption{Diagrammatic representation of {\bf(a)} charged lepton self energy, {\bf (b)} photon self energy with Cutkosky cuts(solid horizontal line)
 for 
 putting particles on mass shell.}
 \label{lep-self}
\end{figure}
where $dt$ is the time of interaction, $dS$ is the differential area, $dl$ and $v(=\frac{|{\bf k}|}{E({\bf k})})$ stands for the length of interaction and velocity, 
respectively and $dV$ is the volume element inside the nucleus. $m_l$ is the lepton mass and
$Im \Sigma(k)$ is the imaginary part of the lepton self energy
(shown in Fig.\ref{lep-self}{\bf (a)})
which is obtained by using the Feynman rules for the lepton self energy($\Sigma(k)$) given by
\begin{equation}\label{defn}
\Sigma (k) = i e^2 \; \int \frac{d^4 q}{(2 \pi)^4} \;
\frac{1}{q^4} \;
\frac{1}{2m_l} \;
L_{\mu \nu} \; \frac{1}{k'^2 - m_l^2 + i \epsilon} \; \Pi^{\mu \nu} (q),
\end{equation}
where $\Pi^{\mu\nu}(q)$ is the photon self energy
which has been shown in Fig.\ref{lep-self}{\bf (b)}.

Using Eq.(\ref{defn}) in Eq.(\ref{defxsec1}), the scattering cross section~\cite{Marco:1995vb} 
is obtained as
\begin{equation}\label{dsigma_3}
\frac {d^2\sigma _A}{d\Omega dE'}=-\frac{\alpha}{q^4}\frac{|\bf{k^\prime}|}{|\bf {k}|}\frac{1}{(2\pi)^2} L_{\mu\nu} \int  Im [\Pi^{\mu\nu}(q)] d^{3}r
\end{equation}
Now comparing Eq.(\ref{eA}) and Eq.(\ref{dsigma_3}), one may write the nuclear hadronic tensor $W_A^{\mu \nu}$ 
in terms of the photon self energy as:
\begin{eqnarray}\label{wamunu}
W_A^{\mu \nu}=-\frac{1}{4\pi^2\alpha} \int  Im [\Pi^{\mu\nu}(q)] d^{3}r
\end{eqnarray}

Using the Feynman rules, the
expression for $\Pi^{\mu\nu}(q)$ is obtained as
\begin{eqnarray}\label{photonse}
\Pi^{\mu \nu} (q)&=& e^2 \int \frac{d^4 p}{(2 \pi)^4} G (p) 
\sum_X \; \sum_{s_p, s_l} {\prod}_{\substack{i = 1}}^{^n} \int \frac{d^4 p'_i}{(2 \pi)^4} \; \prod_{_l} G_l (p'_l)\; \prod_{_j} \; D_j (p'_j)\nonumber \\  
&&  <X | J^{\mu} | H >  <X | J^{\nu} | H >^* (2 \pi)^4  \; \delta^4 (q + p - \sum^n_{i = 1} p'_i),\;\;\;
\end{eqnarray}
 where $G_l$ is the nucleon propagator and $D_j$ is the meson propagator. In the above expression, $<X | J^{\mu} | H >$ is the hadronic current; $s_p$ and $s_l$ are respectively, the spins of 
nucleon and fermions in the final hadronic state $X$. $G (p)$ is the relativistic nucleon propagator inside the nuclear medium which is 
obtained using perturbative expansion of Dyson series in terms of the nucleon self energy($\Sigma^N$) for an interacting Fermi sea. The nucleon self energy
may be obtained using many body field theoretical approach in terms of spectral functions~\cite{FernandezdeCordoba:1991wf, Marco:1995vb}. Therefore, the
nucleon propagator $G (p)$ inside the nuclear medium may also be expressed in terms of the particle and hole spectral
functions as~\cite{FernandezdeCordoba:1991wf}:
\begin{eqnarray}\label{Gp}
G (p) =&& \frac{M_N}{E({\bf p})} 
\sum_r u_r ({\bf p}) \bar{u}_r({\bf p})
\left[\int^{\mu}_{- \infty} d \, \omega 
\frac{S_h (\omega, {\bf{p}})}{p_0 - \omega - i \eta}
+ \int^{\infty}_{\mu} d \, \omega 
\frac{S_p (\omega, {\bf{p}})}{p_0 - \omega + i \eta}\right]\,,
\end{eqnarray}
where $u$ and $\bar u$ are respectively the Dirac spinor and its adjoint, $\mu\left(=\frac{p_F^2}{2 M_N} + Re\left[ \Sigma^N{\tiny \left( \frac{p_F^2}{2 M_N}, p_F \right)} \right]\right)$ is the chemical potential and $p_F$
is the Fermi momentum. $S_h$ and $S_p$, respectively, stand for hole and particle spectral functions, the expression for which is taken from 
 Ref.~\cite{FernandezdeCordoba:1991wf}. The spectral functions contain the
information about the nucleon dynamics in the nuclear medium.
 All the parameters of the spectral function are determined by fitting the binding energy per nucleon and the Baryon number for each nucleus.
 Therefore, we are left with no free parameter. For more discussion please see Ref.~\cite{Haider:2016zrk,Marco:1995vb}.

 To obtain the contribution to the nuclear hadronic tensor $W^{\mu \nu}_{A}$, which is coming from the bound nucleons i.e. $W^{\mu \nu}_{A,N}$, due to the 
 scattering of the charged leptons on the nuclear targets,
 we use Eq.(\ref{photonse}) and Eq.(\ref{Gp}) in Eq.(\ref{wamunu}), and express $W^{\mu \nu}_{A,N}$ 
 in terms of the nucleonic tensor $W^{\mu \nu}_{N}$ convoluted over the hole spectral function $S_h$, and get 
\begin{equation}	\label{conv_WA}
W^{\mu \nu}_{A,N} = 2 \sum_{\tau=p,n} \int \, d^3 r \, \int \frac{d^3 p}{(2 \pi)^3} \, 
\frac{M_N}{E ({\bf p})} \, \int^{\mu_\tau}_{- \infty} d p_0 S_h^\tau (p_0, {\bf p}, \rho^\tau(r))
W^{\mu \nu}_{\tau} (p, q), \,
\end{equation}
 where $\rho^\tau(r)$ is the proton/neutron density inside the nucleus 
  which is determined from the electron-nucleus scattering experiments
 and $S_h^\tau$ is the hole spectral function for the proton/neutron.

 We take the $zz$ component in Eq.(\ref{conv_WA}) for $W_{A,N}^{\mu\nu}$ and $W_{\tau}^{\mu\nu}$, 
the momentum transfer $\bf q$ along the $z$-axis, and using $F_{2N} (x) = \nu W_{2N} (\nu,Q^2)$, we obtain $ F_{{2A}} (x_A,Q^2)$ as~\cite{Haider:2016zrk}:
 \begin{eqnarray} \label{em_f2_noniso}
F_{{2A,N}} (x_A,Q^2)  &=&  2\sum_{\tau=p,n} \int \, d^3 r \, \int \frac{d^3 p}{(2 \pi)^3} \, 
\frac{M_N}{E ({\bf p})} \, \int^{\mu_\tau}_{- \infty} d p_0 S_h^\tau (p_0, {\bf p}, \rho^\tau(r)) \times \left(\frac{M_N}{p_0~-~p_z~\gamma}\right) ~F_{2\tau}(x_N,Q^2)\nonumber \\
&\times&\left[\frac{Q^2}{q_z^2}\left( \frac{|{\bf p}|^2~-~p_{z}^2}{2M_N^2}\right) +  \frac{(p_0~-~p_z~\gamma)^2}{M_N^2} \left(\frac{p_z~Q^2}{(p_0~-~p_z~\gamma) q_0 q_z}~+~1\right)^2\right]. 
\end{eqnarray}
 Similarly, taking the $xx$ component of the nucleon and nuclear hadronic tensors, and using $F_{1N} (x) = M_N W_{1N} (\nu,Q^2)$, we obtain $F_{{1A,N}} (x_A, Q^2)$ as
~\cite{Haider:2016zrk}:
\begin{eqnarray}\label{conv_WA1}
F_{{1A,N}} (x_A, Q^2) &=& 2\sum_{\tau=p,n} AM_N \int \, d^3 r \, \int \frac{d^3 p}{(2 \pi)^3} \, 
\frac{M_N}{E ({\bf p})} \, \int^{\mu_\tau}_{- \infty} d p_0 S_h^\tau (p_0, {\bf p}, \rho^\tau(r)) \times \nonumber\\
&&\left[\frac{F_{1\tau}(x_N, Q^2)}{M_N}+ \frac{{p_x}^2}{M_N^2} \frac{F_{2\tau}(x_N, Q^2)}{\nu_N}\right],\;\;\mbox{where $x_N=\frac{Q^2}{2(p_0q_0 - p_zq_z)}$.}
\end{eqnarray}

 Moreover, in a nucleus, the virtual photon
may interact with the virtual
mesons leading to the modification of the nucleon structure functions due to the additional contributions of the mesons. 
In the numerical calculations we have considered $\pi$ and $\rho$ mesons.
To obtain the contributions of $\pi$ and $\rho$ mesons to the structure functions we follow the similar procedure as in the 
case of nucleon with a difference that the spectral function is now replaced by the dressed meson propagator~\cite{Marco:1995vb, Haider:2016zrk}. We find that
\begin{eqnarray} 
\label{pion_f21}
F_{_{2 A,\pi(\rho)}}(x,Q^2)  &=&  -6\times a \int \, d^3 r \, \int \frac{d^4 p}{(2 \pi)^4} \, 
        \theta (p_0) ~\delta I m D_{\pi(\rho)} (p) \;2m_{\pi(\rho)}~\left(\frac{m_{\pi(\rho)}}{p_0~-~p_z~\gamma}\right)\times \nonumber \\
&&\left[\frac{Q^2}{(q_z)^2}\left( \frac{|{\bf p}|^2~-~(p_{z})^2}{2m_{\pi(\rho)}^2}\right)  
+  \frac{(p_0~-~p_z~\gamma)^2}{m_{\pi(\rho)}^2} \left(\frac{p_z~Q^2}{(p_0~-~p_z~\gamma) q_0 q_z}~+~1\right)^2\right] F_{_{2,\pi(\rho)}}(x_{\pi(\rho)},Q^2),\\
\label{meson_f1}
 F_{_{1A,\pi(\rho)}}(x,Q^2)  &=& -6\times a \times A M_N \int \, d^3 r \, \int \frac{d^4 p}{(2 \pi)^4} \, 
        \theta (p_0) ~\delta I m D_{\pi(\rho)} (p) \;2m_{\pi(\rho)}~\nonumber\\
     &\times&  \left[\frac{F_{1,\pi(\rho)} (x_{\pi(\rho)},Q^2)}{m_{\pi(\rho)}} +\frac{|{\bf p}|^2-p_z^2}{2(p_0q_0-p_zq_z)}\frac{F_{2,\pi(\rho)}(x_{\pi(\rho)},Q^2)}{m_{\pi(\rho)}}\right],~~
\end{eqnarray}
where $x_{\pi(\rho)}=\frac{Q^2}{-2 p \cdot q}$, $m_{\pi(\rho)}$ is the mass of pi(rho) meson and the constant factor $a$ is 1 in the case of pi meson and 2 in the case of $\rho$ meson~\cite{Marco:1995vb}.
$D_{\pi(\rho)}(p)$ is the meson propagator which is given by
\begin{eqnarray}
 D_{\pi(\rho)}(p)=[p_0^2-{\bf p}^2-m_{\pi(\rho)}^2-\Pi_{\pi(\rho)}(p_0,{\bf p})]^{-1},
\end{eqnarray}
where $\Pi_{\pi(\rho)}$ is the meson self energy defined in terms of the form factor $F_{_{\pi(\rho)NN}}(p)$ and irreducible self energy $\Pi_{\pi(\rho)}^\ast$ as
\begin{eqnarray}\label{fpinn}
  \Pi_{\pi(\rho)}&=&\frac{\left(\frac{f^2}{m_\pi^2}\right)~c'_{\pi(\rho)}~F_{_{{\pi(\rho)}NN}}^2(p) {\bf p}^2 \Pi_{\pi(\rho)}^\ast}{1-{f^2 \over m_\pi^2 }V_j^\prime \Pi_{\pi(\rho)}^\ast}\;,\;\mbox{where}\;\;
  F_{_{{\pi(\rho)}NN}}(p) = \left(\frac{\Lambda^2 - m_{\pi(\rho)}^2}{\Lambda^2 + {\bf p}^2}\right).
\end{eqnarray}
In the above expression, $V_j^\prime=V_L^\prime$($V_T^\prime$) for the pi(rho) meson, 
are the longitudinal(transverse) part of spin-isospin interaction, respectively, the
expressions for which are taken from the Ref.~\cite{Marco:1995vb} with
$c'_{\pi}=1$ and $c'_{\rho}=3.94$, $\Lambda=1~GeV$ and $f=1.01$. 
 These parameters have been fixed in our earlier works~\cite{Haider:2016zrk, Haider:2015}
while describing nuclear medium effects in the electromagnetic nuclear structure function $F_{2A} (x,Q^2)$ to explain the 
latest data from the JLab and other experiments performed using charged lepton scattering from several nuclear targets in the DIS region.

We now define the total EM nuclear structure functions $F_{iA} (x,Q^2)$(i=1,2) which include the nuclear medium effects as:
\begin{equation}\label{ftotal}
 F_{iA} (x,Q^2) =  F_{iA,N} (x,Q^2) +  F_{iA,\pi} (x,Q^2) + F_{iA,\rho} (x,Q^2)\;;\;\; i=1,2.
\end{equation}
We, therefore, define $F_{LA} (x,Q^2)$, $R_{A} (x,Q^2)$ and $R_{2A} (x,Q^2)$ 
in nuclear targets in analogy with $F_{LN} (x,Q^2)$, $R_N(x,Q^2)$ and $R_{2N} (x,Q^2)$ as:
\begin{equation}\label{fla}
 F_{LA} (x,Q^2) = \left( 1+ {4 M_N^2 x^2 \over Q^2}\right)F_{2A} (x,Q^2) - 2 x F_{1A} (x,Q^2),
\end{equation}

\begin{equation}\label{rlar2a}
 R_{A} (x,Q^2)={F_{LA} (x,Q^2) \over 2 x F_{1A} (x,Q^2)} = \frac{\sigma_{LA}(x,Q^2)}{\sigma_{TA}(x,Q^2)} =\left( 1+ \frac{4 M_N^2 x^2}{Q^2}\right) R_{2A} (x,Q^2) -1,\;\;
 \mbox{where} \;\;\; R_{2A} (x,Q^2)= {F_{2A} (x,Q^2) \over 2 x F_{1A} (x,Q^2)}.
\end{equation}
\begin{figure}[t]
\begin{center}
\includegraphics[height= 7cm , width= 0.85\textwidth]{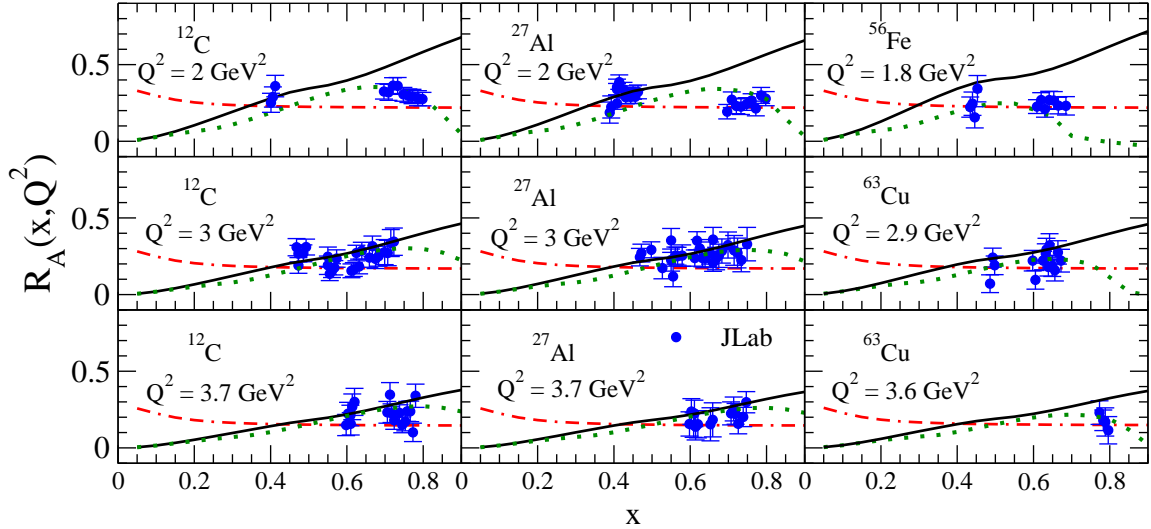}
\end{center}
\caption{$R_{A} (x,Q^2)=\frac{F_{LA} (x,Q^2)}{2 x F_{1A} (x,Q^2)}$ ($A = ^{12}C$, $^{27}Al$ (isoscalar), $^{56}Fe, 
 ~^{63}Cu$ (nonisoscalar)) vs $x$ are shown at different $Q^2$. The results are obtained using the present model {\bf(i)} without any kinematical cut on CM energy(solid line) and
 {\bf (ii)} with a kinematical cut on CM energy $W > 1.4~GeV$(dotted line).
 The results are compared with the results for the
 free nucleon case obtained by using the parameterization of Whitlow et al.~\cite{Whitlow:1991uw}(double dashed-dotted line) and the 
 experimental data of the JLab(bold circles)~\cite{Mamyan:2012th}.}
\label{plb:fig1}
\end{figure}

\begin{figure}[t]
\begin{center}
\includegraphics[height= 7cm , width= 0.85\textwidth]{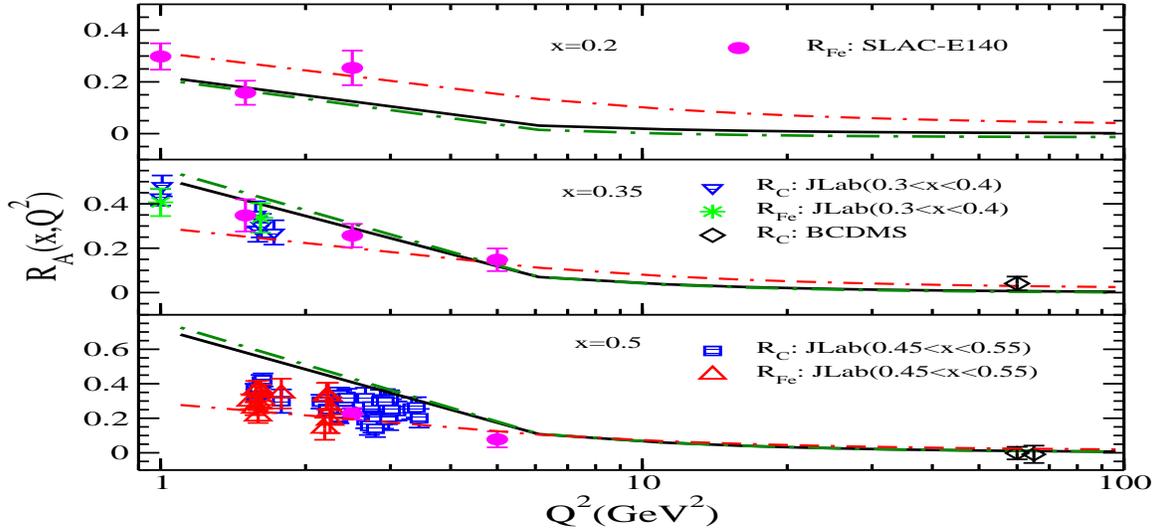}
\end{center}
\caption{$R_{A} (x,Q^2)=\frac{F_{LA} (x,Q^2)}{2 x F_{1A} (x,Q^2)}$ ($A = ^{12}C$ and $^{56}Fe$) vs $Q^2$ are shown at different $x$ obtained by using the present model for isoscalar nuclear targets. 
Numerical results in $^{12}C$(solid line) and in $^{56}Fe$(dashed-dotted line)
  are compared with the results of free nucleon case(double dashed-dotted line) obtained by using the parameterization of Whitlow et al.~\cite{Whitlow:1991uw}. 
  The results are also compared with the experimental data from the JLab(in $^{12}C$ and $^{56}Fe$)~\cite{Mamyan:2012th}, SLAC-E140(in $^{56}Fe$)~\cite{Dasu:1988ms} and 
 the BCDMS(in $^{12}C$)~\cite{Benvenuti:1989rh}.}
\label{plb:fig2}
\end{figure}
\section{Results}\label{sec_results}
 The numerical results for 
 $R_{A} (x,Q^2)$ defined in Eq.(\ref{rlar2a}) are calculated using $F_{iA,N} (x,Q^2)$, $F_{iA,\pi(\rho)} (x,Q^2)\;\;(i=1,2)$ given in Eqs.(\ref{em_f2_noniso}), (\ref{conv_WA1}),  
 (\ref{pion_f21}) and (\ref{meson_f1}).
  The quark/antiquark PDFs of the nucleon 
 have been taken from 
 the parameterization given by the CTEQ collaboration~\cite{cteq} for the calculations of $F_{2N} (x,Q^2)$ which are performed in four flavor($u,~d,~s,$ and $c$) scheme 
 at the next-to-leading order(NLO)~\cite{Vermaseren, Neerven}. $F_{1N} (x,Q^2)$ has been calculated using $F_{2N} (x,Q^2)$ given in Eq.(\ref{eq:f22xf1}) and $R_{N} (x,Q^2)$ 
 given by Whitlow et al.~\cite{Whitlow:1991uw}. We have incorporated the target mass correction (TMC) at the nucleon level
 following Ref.~\cite{Schienbein:2007gr}. 
For the pions, we have taken the parton distribution functions given by 
 Gluck et al.~\cite{Gluck:1991ey} and for the rho mesons, we have used the same PDFs as for the pions.  
 We have also considered the effect of 
 shadowing following Ref.~\cite{Kulagin:2004ie}. 

  We present the theoretical results for $R_{A}(x,Q^2)$ in Fig.\ref{plb:fig1} for $^{12}C$, $^{27}Al$, $^{56}Fe$ and $^{63}Cu$ and compare them with the 
  recent experimental results from the JLab~\cite{Mamyan:2012th}. The results are also compared with the free nucleon case i.e. $R_{N}(x,Q^2)$ as 
  parameterized by Whitlow et al.~\cite{Whitlow:1991uw}.   
  The results have been presented for the two cases
 (i) without applying any cut on the CM energy $W$(solid line), thus including the contributions to $R_{A}(x,Q^2)$ from the allowed kinematic region in $W$ and 
 (ii) by applying a cut on $W \ge 1.4~ GeV$(dotted line), the case in which $R_{A}(x,Q^2)$ does not include the contribution from the $\Delta(1232)$ resonance region. 

 We observe that: 
 \begin{enumerate}
  \item  Our treatment of the nuclear medium effects without a cut on $W$(solid line) satisfactorily reproduces the experimental data for $R_A(x,Q^2)$ from the JLab~\cite{Mamyan:2012th} for $Q^2 \ge 2$ $GeV^2$
but overestimates them at lower $Q^2$($Q^2 \le 2$ $GeV^2$) in the region of $x \ge 0.5$,
\item Our theoretical results with a cut on $W\ge 1.4$ $GeV$(dotted line) satisfactorily explain the experimental data in the 
entire region of $Q^2$($1 < Q^2<5$ $GeV^2$) and $x$($0.5 < x < 0.8$) of the JLab experiment~\cite{Mamyan:2012th}.
\item The nuclear medium effects on $R_A(x,Q^2)$ show a structure with $x$, i.e. at lower $x$($x<x_{o}$) the nuclear medium effect decreases the value 
 of $R_A(x,Q^2)$ i.e. $R_A(x,Q^2)< R_N(x,Q^2)$
while for higher $x$($x>x_{o}$) it increases i.e. $R_A(x,Q^2)\ge R_N(x,Q^2)$ with a cross over at $x=x_{o}$. The value of $x_{o}$ is 
 dependent on $Q^2$ and lies between 0.35 and 0.55 for $2<Q^2<10$ $GeV^2$. In the kinematic region of the JLab data~\cite{Mamyan:2012th} shown in Fig.\ref{plb:fig1}, i.e. $2<Q^2<3.7$ $GeV^2$
 and $x \ge 0.4$, the nuclear medium effects are small
without the inclusion of first resonance region($W~>~1.4$ $GeV$) but are enhanced when the contribution of the first resonance region 
 (without cut on W) is also included. 
 In both the cases (with or without cut on W), the 
nuclear medium effects decrease with $Q^2$ and have almost no dependence on the mass number $A$. The above observations imply that: 
\begin{enumerate}
 \item In the region of $Q^2\le2$ $GeV^2$, our results of $R_A(x,Q^2)$ 
 are larger than the experimental data. These results are obtained using the DIS formalism in the entire kinematic region which includes the 
 first resonance region(i.e. 1.3 $GeV^2 < W^2 < 1.9$ $GeV^2$) dominated by $\Delta(1232)$ and possibly
 $N^*(1440)$ resonance. A comparison of our results with and without cut of $W$ suggests that in the kinematic  
region of $Q^2 < 2~ GeV^2$, the use of DIS formalism is not appropriate
and a calculation of $R_A(x,Q^2)$, using explicit mechanism for $\Delta(1232)$ and $N^\ast(1440)$ resonance excitations 
 in the nuclear medium would give a better description of $R_A(x,Q^2)$.
\item The contribution to $R_A(x,Q^2)$ from the resonance region is also well reproduced by the DIS formalism at high $Q^2$.
This supports the conclusion drawn from the study of Bloom-Gilman duality in the free nucleon case, that 
 the quark-hadron(QH) duality works
better in the higher resonance region than in the region of $\Delta(1232)$ resonance~\cite{Melnitchouk:2005zr} as our results for $R_A(x,Q^2)$ using DIS formalism seem to agree better with the experimental
results in the higher $Q^2$ region than in the lower $Q^2$ region specially for $x\ge 0.5$.
\end{enumerate}
 \begin{figure}
\begin{center}
\includegraphics[height= 7cm , width= 0.85\textwidth]{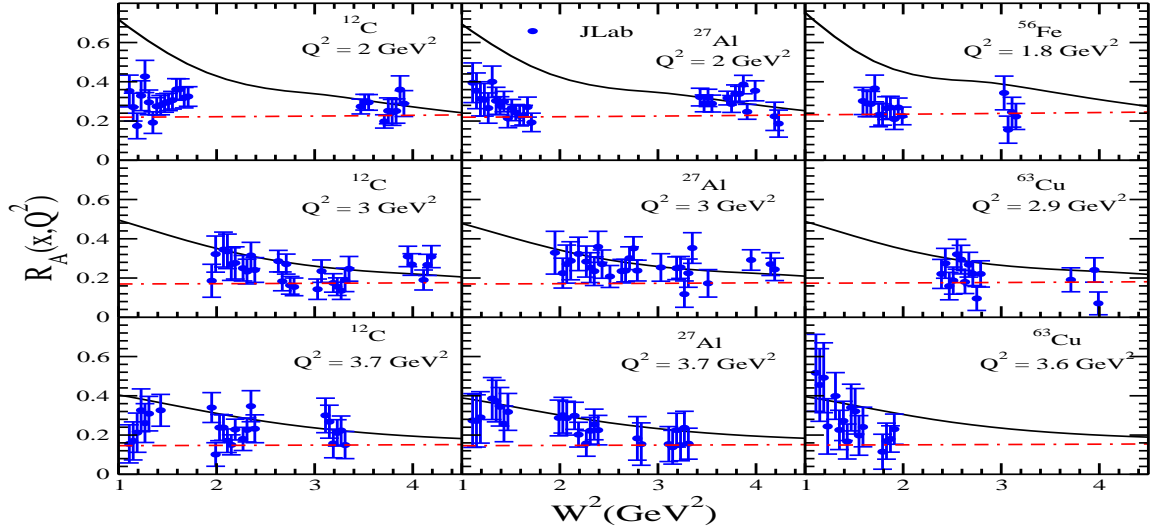}
\end{center}
\caption{Results for $R_{A} (x,Q^2)=\frac{F_{LA} (x,Q^2)}{2 x F_{1A} (x,Q^2)}$($A=$ $^{12}C$, $^{27}Al$ (isoscalar), $^{56}Fe$, $^{63}Cu$ (nonisoscalar)) vs $W^2$ are shown at different $Q^2$. Numerical 
results obtained using the present model(solid line) are compared with the results of free nucleon case using the parameterization of Whitlow et al.~\cite{Whitlow:1991uw}(double dashed-dotted line) and with the 
 experimental data of the JLab~\cite{Mamyan:2012th}(bold circles).}
\label{plb:fig4}
\end{figure}

\begin{figure}
\begin{center}
\includegraphics[height= 5.0cm , width= 0.8\textwidth]{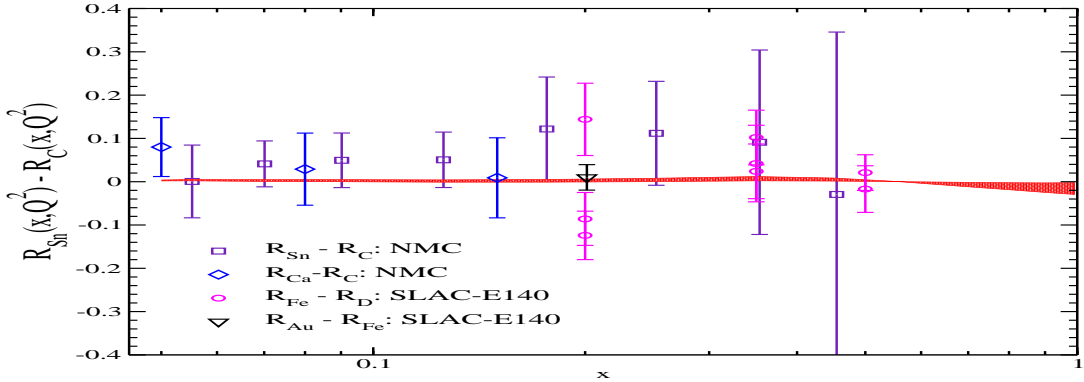}
\end{center}
\caption{Numerical results for $R_{Sn}(x,Q^2) - R_{C}(x,Q^2)$ vs $x$ are obtained using the 
present model for $4<Q^2<35$ $GeV^2$(narrow band) and the nuclear targets are treated as isoscalar. These results are compared with the experimental
results for $R_{Fe}(x,Q^2) - R_{D}(x,Q^2)$~\cite{Dasu:1988ms}(circle), $R_{Au}(x,Q^2) - R_{Fe}(x,Q^2)$~\cite{Dasu:1988ms}(down triangle), $R_{Ca}(x,Q^2) - R_{C}(x,Q^2)$~\cite{Amaudruz:1992wn}(diamond)
 and $R_{Sn}(x,Q^2) - R_{C}(x,Q^2)$~\cite{Arneodo:1996ru}(square).  }
\label{plb:fig3}
\end{figure}

\begin{figure}[t]
\begin{center}
\includegraphics[height= 4.8cm , width= 0.8\textwidth]{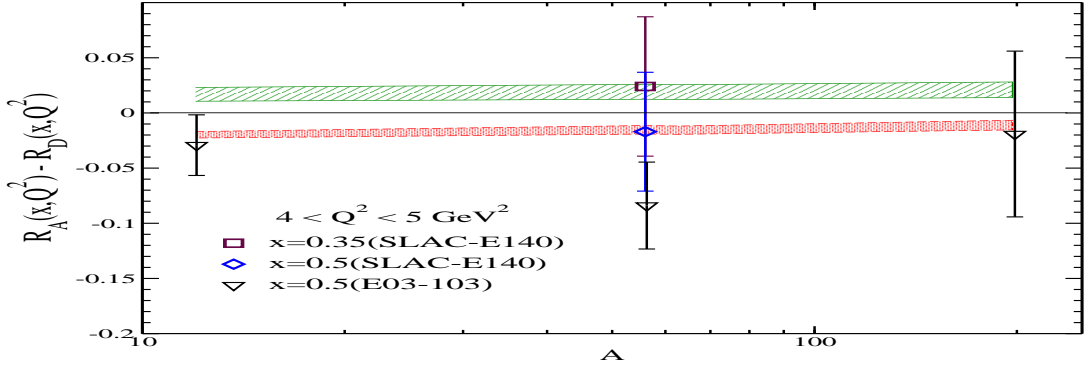}
\end{center}
\caption{Results for $R_{A}(x,Q^2)- R_{D}(x,Q^2)$ vs $A$ obtained using the present model 
 are presented for $^{12}C$, $^{56}Fe$, $^{63}Cu$ and $^{197}Au$ in the range of $4<Q^2<5$ $GeV^2$ 
 for $x=0.4$(lower band) and $x=0.5$(upper band).
 The results are compared with the experimental results of the JLab~\cite{Solvignon:2009it} 
  and the SLAC-E140~\cite{Dasu:1988ms} corresponding to the $x$ values shown in the figure. Nuclear targets are treated as isoscalar.}
\label{plb:fig7}
\end{figure}

\begin{figure}[t]
\begin{center}
\includegraphics[height= 7.0cm , width= 0.9\textwidth]{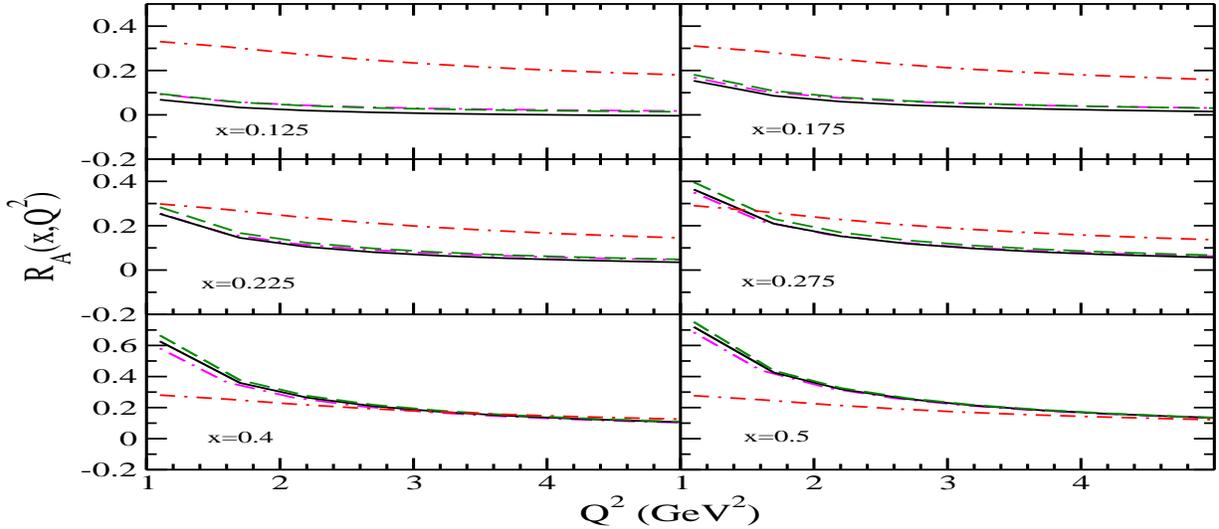}
\end{center}
\caption{Results for $R_{A}(x,Q^2)$ vs $Q^2$ obtained using the present model are shown at different $x$ for some nuclei viz. $^{12}C$(dashed-dotted line), $^{63}Cu$(solid line) and $^{197}Au$(dashed line). 
 The results are compared with the results of
 free nucleon case obtained by using the parameterization of Whitlow et al.~\cite{Whitlow:1991uw}(double dashed-dotted line). Nuclear targets are treated as isoscalar.}
\label{plb:fig5}
\end{figure}
Our theoretical results are also in agreement with the results of $R_A(x,Q^2)$ reported from the SLAC E140~\cite{Dasu:1988ms} and BCDMS~\cite{Benvenuti:1989rh} experiments as shown in Fig.\ref{plb:fig2}
 at the different values of $x$ and $Q^2$ in $^{12}C$ and $^{56}Fe$. 
From the figure it may be observed that the present theoretical results are in agreement with the results of 
 BCDMS experiment~\cite{Benvenuti:1989rh} at high $Q^2$ but overestimate the SLAC E140 results~\cite{Dasu:1988ms} at lower $Q^2$ for
 $x\ge 0.5$. These results also show that the CGR is satisfied in the case of nuclei i.e. $R_A(Q^2) \rightarrow 0$ at high $Q^2$ at the same level of precision as in 
 the case of free nucleon for $0.2 \le x \le 0.5$. At lower $Q^2$, the deviation from the free nucleon case, i.e. $R_N(x,Q^2)$ is in the right direction to explain the experimental data but
 overestimate them.

In order to further discuss the results presented in Fig.\ref{plb:fig1} and Fig.\ref{plb:fig2} above, we have shown in Fig.\ref{plb:fig4}, the values of $R_A(x,Q^2)$
vs $W^2$ for the various values of $Q^2$. We find that at higher $Q^2$($Q^2 > 2$ $GeV^2$),  our results are in fair agreement with the experimental data for $W^2 \le 4$ $GeV^2$ obtained from the JLab~\cite{Mamyan:2012th} 
in $^{12}C$, $^{27}Al$, $^{56}Fe$, and $^{63}Cu$. However, at lower $Q^2$($Q^2 \le 2$ $GeV^2$), the theoretical results using DIS formalism overestimate $R_A(x,Q^2)$
for $W^2 <2.5$ $GeV^2$ which describes the resonance region dominated by the $\Delta(1232)$ and $N^\ast(1440)$ resonances. 
 This strengthens our assertion that in the region of low $Q^2$ and $W^2$ 
a realistic calculation of $R_A(x,Q^2)$ using explicitly the 
$\Delta(1232)$ and $N^\ast(1440)$ resonance excitations in the nuclear medium should be more appropriate than the use of DIS formalism. The theoretical results obtained using the 
present model for $R_A(x,Q^2)$ for all the nuclei shown in Fig.\ref{plb:fig4}
are in better agreement with the experimental results when the nuclear medium effects are included 
specially in the region of $W^2 \ge 2.5$ $GeV^2$ for all $Q^2$ considered here.
 We also observe that the effect of nuclear medium leads to an enhancement in the value of $R_A(x,Q^2)$ over $R_N(x,Q^2)$ in the region of $W^2$ considered here. 
 This enhancement is quite small at 
 higher $W^2$ but is large at lower $W^2$. It decreases with the increase in $Q^2$ and is almost independent of $A$. 

 The world data on nuclear dependence of $R_A(x,Q^2)$ have been compiled and reanalyzed by Guzey et al.~\cite{Guzey:2012yk} and Solvignon et al.~\cite{Solvignon:2009it}.
We show in Fig.\ref{plb:fig3}, a comparison of our present results with the experimental data summarized by Guzey et al.~\cite{Guzey:2012yk}. 
 A reasonable agreement with the experimental data~\cite{Arneodo:1996ru, Amaudruz:1992wn, Dasu:1988ms} is seen for 
$R_A(x,Q^2) - R_{A^\prime}(x,Q^2)$ for $A$ and $A^\prime$ nuclei as shown in Fig.\ref{plb:fig3}. Our results
are also in agreement with the recent experimental results from the JLab~\cite{Mamyan:2012th} for $R_{A}(x,Q^2) - R_{C}(x,Q^2)$ for $A=~^{27}Al$ and $^{63}Cu$ which are not reproduced here as they
 have very large uncertainties due to systematic and statistical errors. We also show the $A$ dependence of $R_A(x,Q^2) - R_D(x,Q^2)$ in Fig.\ref{plb:fig7} and 
 compare it with the data of JLab~\cite{Solvignon:2009it} and SLAC~\cite{Dasu:1988ms}. We find that $R_A(x,Q^2) - R_D(x,Q^2)$
 is negative for $x<0.45$ and changes sign for $x>0.45$ and it has almost no dependence on $A$ in the absence of Coulomb effects which are not considered in this work.
 
A precise determination of the structure functions in nuclei using the method of Rosenbluth separation requires a knowledge of
$R_A(x,Q^2)$ with high accuracy. An experiment to measure the nuclear dependence of $R_A(x,Q^2)$ and to determine the structure
functions $F_{1A}(x,Q^2)$, $F_{2A}(x,Q^2)$ and $F_{LA}(x,Q^2)$ with high precision in $^{12}C$, $^{63}Cu$ and $^{197}Au$ nuclei
has been proposed at the JLab~\cite{Covrig} in the kinematic region of $0.1<x<0.6$ and $1<Q^2<5$ $GeV^2$ for $W^2$ up to
10 $GeV^2$. We, therefore, show in the Fig.\ref{plb:fig5} the results of $R_A(x,Q^2)$ vs $Q^2$  for various values of $x$ 
 using the present model relevant for the kinematics of the JLab experiment~\cite{Covrig}. 
 
 A comparison of our results with the experimental results of the JLab experiment~\cite{Covrig} in the future will help to obtain a more precise determination of the nuclear structure 
functions and provide further insight in understanding the physics of the nuclear medium effects in the DIS of the electrons from nuclear targets.
 \end{enumerate}
\section{Summary}\label{sec_summary}
 We have in this work studied the nuclear medium effects on the ratio 
 $R_A(x,Q^2)= \frac{F_{LA} (x,Q^2)}{2xF_{1A} (x,Q^2)}$
 and its impact on the Callan-Gross relation(CGR) in nuclei, using a microscopic nuclear model and considered the effects of 
 the Fermi motion, binding energy, nucleon correlations, mesonic contribution and shadowing.
 We find that
  \begin{itemize}
 \item The inclusion of nuclear medium effects leads to a better description of the experimental data from JLab\cite{Mamyan:2012th}, 
 SLAC~\cite{Dasu:1988ms}, BCDMS~\cite{Benvenuti:1989rh} and NMC~\cite{Amaudruz:1992wn, Arneodo:1996ru} in various nuclei in a wide range 
 of $x$ and $Q^2$. At high $Q^2$ the experimental results are well reproduced, while at low $Q^2$($\le~ 2 ~GeV^2$) we overestimate the experimental data for $x \ge 0.5$.
 
\item In nuclei, the Callan-Gross relation is satisfied at high $Q^2$ with almost the same precision as in the case of free nucleon for all the values of $x$.
 At lower and moderate $Q^2$, there is deviation in $R_A(x,Q^2)$ from the free nucleon value due to the nuclear medium effects and it is in the right direction 
 to give a better description  of the available experimental data but overestimates them for $x > 0.2$.
 
 \item With the inclusion of nuclear medium effects, $R_A(x,Q^2)$ is almost independent of $A$ for $A \ge$ 12 subject to the Coulomb corrections which are small and relevant only 
 for the DIS of the low energy electrons from high Z nuclear targets.

 \item The use of DIS formalism to calculate the contribution of $R_A(x,Q^2)$ in the region of low $W^2$ and low $Q^2$ overestimates the experimental 
 results in this region. In this
 kinematic region an explicit calculation of $R_A(x,Q^2)$ arising due to the resonance excitation of $\Delta(1232)$ and $N^*$(1440) in the nuclear medium should be more appropriate. This 
 also indicates that the phenomenon of QH duality in nuclei works better in the kinematic region beyond the first resonance region as in the case of the free nucleon.
 \end{itemize}

\end{document}